\documentclass[preprint,amsmath,amssymb,prb]{revtex4}
\usepackage{graphicx}
\usepackage{amssymb}
\usepackage{epsfig}
\countdef\preprint=0 \countdef\tube=12
\begin{document}

\bibliographystyle{aip}
%**end of header

\title{\underline{Sine-Gordon Model}\\

Renormalization Group Solution and Applications}

\author{Mariana Malard}
\affiliation{Faculdade UnB Planaltina, Universidade de Brasilia,
73300-000 Planaltina-DF, Brazil}

\date{February, 2012}

\begin{abstract}
The sine-Gordon model is discussed and analyzed within the framework
of the renormalization group theory. A perturbative renormalization
group procedure is carried out through a decomposition of the
sine-Gordon field in slow and fast modes. An effective slow modes's
theory is derived and re-scaled to obtain the model's flow
equations. The resulting Kosterlitz-Thouless phase diagram is
obtained and discussed in detail. The theory's gap is estimated in
terms of the sine-Gordon model paramaters. The mapping between the
sine-Gordon model and models for interacting electrons in one
dimension, such as the g-ology model and Hubbard model, is discussed
and the previous renormalization group results, obtained for the
sine-Gordon model, are thus borrowed to describe different aspects
of Luttinger liquid systems, such as the nature of its excitations
and phase transitions. The calculations are carried out in a
thorough and pedagogical manner, aiming the reader with no previous
experience with the sine-Gordon model or the renormalization group
approach.
\end{abstract}

\maketitle

\large \tableofcontents

\newpage

\section{Introduction}

The sine-Gordon model was originally proposed as a toy model for
interacting quantum field theories and it has been intensively
investigated ever since.

In low dimensional physics, the sine-Gordon model often appears as a
description of systems with non-quadratic interactions having a
strong pining effect. Contrary to the quadratic momentum that
promotes fluctuations in the system, the sine-Gordon potential would
like to lock the model field in one of the minima of the cosine. The
model is particularly useful to describe strongly correlated
electronic systems in one dimension.

As it is well know, interacting electrons systems in dimensions
higher than $D=1$ are well described by Landau's Fermi liquid
theory. In $D=1$, however, the Fermi liquid fails due to an
instability - known as Peierls instability - generated by $2k_{F}$
scattering processes which are particular to one-dimensional Fermi
``surfaces". In opposition to the Fermi liquid nomenclature,
one-dimensional electronic systems are generically referred to as
Luttinger liquids, after the related work by Luttinger.
\cite{Luttinger}

Throughout the years, many models and formalisms have been proposed
to describe the special behavior of Luttinger liquids. Of particular
interest is the bosonization mapping between different Luttinger
liquid hamiltonians, such as the $g$-ology model and Hubbard models,
and the sine-Gordon model for which plenty of anallitic results are
available.

\newpage

The S-matrix formalism for the sine-Gordon model and the elementary
excitations spectrum have been analytically derived
\cite{Karowski,Zamolodchikov,Faddeev,Korepin}. Results for the
sine-Gordon model form factors \cite{Babujian} and finite size
correction to the model's spectrum \cite{Destri,Fioravanti,Feverati}
are also available in the literature.

Nevertheless, there still are a number of open questions regarding
the sine-Gordon model. Exact results for the model's correlation
functions are still lacking, for example. The evaluation of the
model's spectrum has proved challenging as well \cite{Niccoli}.

%\footnote{To this %date, the best results for the sine-Gordon model spectrum are based
%on the Bethe ansatz and, thus, have problems related to the
%completeness of the the set of states generated by the method.}

The renormalization group theory is an important analitic tool is
this context for it provides the understanding of the sine-Gordon
model's phase transition and the energy scale at which it occurs.

\newpage

\section{The Model}

The sine-Gordon model model hamiltonian is given by
\begin{equation}
H[\Pi,\varphi]=\int
dx\,\left[\,\frac{v}{2}(\Pi^{2}+(\partial_{x}\varphi)^{2})-\tilde{g}\cos(\beta\varphi)\,\right]
\label{H}
\end{equation}
where $\varphi=\varphi(x,t)$ and $\Pi=\Pi(x,t)$ are canonically
conjugated fields, that is:
\begin{equation}
\partial_{t}\varphi(x,t)=v\Pi(x,t)
\label{relfields}
\end{equation}

The model lagrangean writes:
\begin{equation}
\nonumber L[\varphi]=\int dx\,\{\,\partial_{t}\varphi\,.\,\Pi\,-\,H[\Pi,\varphi]\,\}
\end{equation}
\begin{equation}
L[\varphi]=\int
dx\,\left[\,\frac{1}{2v}(\partial_{t}\varphi)^{2}-\frac{v}{2}(\partial_{x}\varphi)^{2}+\tilde{g}\cos(\beta\varphi)\,\right]
\label{L}
\end{equation}

After an integration by parts, the action
\begin{equation}
\nonumber S[\varphi]=\int dt\,L[\varphi]
\end{equation}
can be written in the following form
\begin{equation}
\nonumber S[\varphi]=\int\int
dtdx\,\left[\,\frac{v}{2}\varphi\partial_{x}^{2}\varphi-\frac{1}{2v}\varphi\partial_{t}^{2}\varphi+\tilde{g}\cos(\beta\varphi)\,\right]
\end{equation}
where we have assumed that the $\varphi$-field goes to zero at the
boundaries of the integration plane.

For a reason that will become clear in Section III.B, it is more
convenient to work in imaginary time $t\rightarrow -it$, i.e., with
the euclidean action $iS\rightarrow S$ which reads
\begin{equation}
S[\varphi]=S_{0}[\varphi]+S_{I}[\varphi]
\label{S}
\end{equation}
\begin{equation}
S_{0}[\varphi]=\int dx\,\frac{1}{2}\varphi\nabla_{x}^{2}\varphi\quad
with\quad\nabla_{x}^{2}=\partial_{x}^{2}+\frac{1}{v^{2}}\partial_{t}^{2}
\label{S0}
\end{equation}
\begin{equation}
S_{I}[\varphi]=\int dx\,l_{I}[\varphi]\quad with\quad
l_{I}[\varphi]=g\cos(\beta\varphi) \label{SI}
\end{equation}
where $x\rightarrow(x,vt)$ and $g=\tilde{g}/v$.

\newpage

\section{Renormalization Group Treatment}

\subsection{Conceptual overview on renormalization group theory}

The renormalization group (R.G.) is essentially a theory of scale
invariance and symmetries. A symmetry or scale operation on a system
is a transformation that portraits the system's appearance and
behavior at different scales (where this scale might be a length
scale, energy, or any typical scale in the system). The system is
said to be scale invariant under a certain scale transformation when
it looks the same at all scales.

In the R.G. theory, the system's microscopic physical quantities
(such as mass, charge and interaction parameters) do depend on the
scale at which the system is observed. The scale invariant
properties are the ones that emerge from the system's macroscopic
structure and are related to the degree of order in the system. The
global invariant properties are represented by physical quantities
called order parameters.

Many states of matter are characterized in terms of scale invariant
properties and order parameters, e.g.: the crystalline lattice
structure in a solid that translates into a periodic density of
particles, the spins orientation in a ferromagnetic material that
results in a net spontaneous magnetization, the localization of
charge in an insulator described in terms of charge density waves,
and so on.

When, for some reason, a scale invariant system loses its invariance
it is said to have undergone a phase transition. The phase
transition, i.e. the loss of invariance, takes place at a certain
critical scale that is typical of each system. The critical scale
defines an energy, called gap (or mass, in quantum field theory
language), that measures the extent of the disturbance in the
system's order parameter. The phase transition's critical point is
defined by the values of the system's parameters at the critical
scale.

As an example, picture a perfect solid at $T=0$. As the temperature
is increased, the solid will eventually lose distance invariance at
some critical length scale that is set by the characteristics of the
material. At this scale, the lattice correlations cannot compete
with the thermal fluctuations and the solid structure melts in a
fluid.

Here, we are interested in the scale behavior of the sine-Gordon
model (or rather of a certain physical system that can be described
by the sine-Gordon model).

The next three sections feature a general presentation of the R.G.
procedure. In Sec. B, the decomposition of a generic quantum field
theory in slow and fast modes is presented; The goal of Sec. C is to
express the so-called residual action that mixes slow and fast modes
in terms of the theory Green's function; In Sec. D, a slow modes'
effective action is derived through averaging out the fast modes.
The last two section are dedicated to the application of the general
formalism to the sine-Gordon model. In Sec. E, the model's effective
action for the slow modes is evaluated. This effective theory is
then renormalized, resulting in a re-scaled sine-Gordon model. The
model's flow equation are derived in Sec. F.

\newpage

\subsection{General procedure I - Decomposition in slow and fast modes}

The R.G. procedure as it is presented in this section follows the
formulation by Kenneth Wilson, developed in the late 60's and which
awarded him the Nobel Prize in 1982. Nowadays, this formulation is
routinely called ``wilsonian approach" to the R.G theory.

The procedure is based on splitting the theory's field $\varphi(x)$
in two components corresponding to different momentum-frequency
regions of the original field's Fourier decomposition.
Mathematically,
\begin{equation}
\nonumber \varphi(x)=\int\frac{dq}{(2\pi)^2}\,\varphi(q)e^{iqx}
\end{equation}
\begin{equation}
\nonumber \varphi(x)=\int_{bulk}\frac{dq}{(2\pi)^2}\,\varphi(q)e^{iqx}+\int_{shell}\frac{dq}{(2\pi)^2}\,\varphi(q)e^{iqx}
\end{equation}

\vspace{.3cm}

where $x\rightarrow(x,vt)$, $q\rightarrow(q,\omega/v)$,
$qx\rightarrow(qx+\omega t)$, $|q|^{2}\rightarrow
q^{2}+\omega^{2}/v^{2}$ and
\begin{equation}
\nonumber bulk\equiv|q|<\frac{\Lambda}{s},
\end{equation}
\begin{equation}
\nonumber shell\equiv\frac{\Lambda}{s}<|q|<\Lambda,
\end{equation}
with $s\approx1$ and where $\Lambda$ is a momentum-frequency cutoff.

Note that, in the original covariant space, the momentum-frequency
shell would correspond to the unbounded surface between the two
hyperbolaes $q^{2}-\omega^{2}/v^{2}=\frac{\Lambda^{2}}{s^{2}}$ and
$q^{2}-\omega^{2}/v^{2}=\Lambda^{2}$. Although this surface imposes
a cutoff in the modulus $|q|$, individually the coordinates $q$ and
$\omega$ remain boundless. Therefore, in the original covariant
space, the integration of a function $f(q,\omega)$ over the shell
will diverge if $f(q,\omega)$ does not decay fast enough. The
purpose of the imaginary time rotation performed before Eq.
(\ref{S}) is exactly to avoid complications that might arise from an
unbounded shell.

In a more compact form, we may write the $\varphi$-field as
\begin{equation}
\varphi(x)=\varphi^{s}(x)+\delta\varphi(x)
\label{phi}
\end{equation}
where:
\begin{equation}
\varphi^{s}(x)=\int_{bulk}\frac{dq}{(2\pi)^{2}}\,\varphi(q)e^{iqx}
\label{phis}
\end{equation}
\begin{equation}
\delta\varphi(x)=\int_{shell}\frac{dq}{(2\pi)^{2}}\,\varphi(q)e^{iqx}
\label{deltaphi}
\end{equation}

\vspace{.3cm}

The $\varphi^{s}$-field contains the so-called slow modes of the
original $\varphi$-field while the $\delta\varphi$-field contains
the fast modes.

The idea is to obtain the theory's action, written for the
$\varphi$-field in the full momentum-frequency space, in terms of
slow and fast mode fields and take its average with respect to the
fast modes' unperturbed ground state. The result of this average is
an effective action for the slow modes. A ``renormalized" theory is
thus obtained from the effective one through a scale transformation,
or renormalization, of the momentum-frequency cutoff. The R.G.
statement, based on the assumed scale invariance of the theory, is
that the original and renormalized theories are equal, i.e. that the
slow modes' effective theory defined in the bulk is equivalent to a
scale renormalization of the full original theory in the entire
momentum-frequency space. This equivalence allows the derivation of
the theory's R.G. flow equations which comprise the final outcome of
the R.G. approach.

Let us proceed by rewriting the action $S[\varphi]$ in terms of the
$\varphi^{s}$- and $\delta\varphi$-fields.

\newpage

Since
\begin{equation}
\nonumber \int dx\,\delta\varphi\nabla_{x}^{2}\varphi^{s}=\int
dx\,\varphi^{s}\nabla_{x}^{2}\delta\varphi=
\end{equation}
\begin{equation}
\nonumber =\int
dx\int_{bulk}\frac{dq}{(2\pi)^{2}}\int_{shell}\frac{dq'}{(2\pi)^{2}}\,\,\varphi(q)\left(-q'^{2}-\frac{\omega'^{2}}{v^{2}}\right)\varphi(q')e^{i(q+q')x}=
\end{equation}

\begin{equation}
\nonumber
=\int_{bulk}\frac{dq}{(2\pi)^{2}}\,\,\varphi(q)\left(-q^{2}-\frac{\omega^{2}}{v^{2}}\right)\varphi(-q)\,\Theta(|q|-\frac{\Lambda}{s})=0
\end{equation}

\vspace{.5cm}

inserting Eq. (\ref{phi}) into Eq. (\ref{S0}) gives:
\begin{equation}
S_{0}[\varphi]=S_{0}[\varphi^{s}]+S_{0}[\delta\varphi]
\label{S01}
\end{equation}

The interaction contribution to the action, that is $S_{I}[\varphi]$
in Eq. (\ref{SI}), will be treated via perturbation theory around
the slow modes $\varphi^{s}$-field. (This procedure is similar to
the usual saddle-point expansion around a fixed classical field
configuration.) Up to second order in the perturbation, i.e. to
second order in the fast modes $\delta\varphi$-field, we have
\begin{equation}
S_{I}[\varphi]=S_{I}[\varphi^{s}]+\int
dx\,a^{s}(x)\delta\varphi(x)+\int
dxdx'\,\delta\varphi(x)b^{s}(x,x')\delta\varphi(x') \label{SI1}
\end{equation}
where the coefficients $a^{s}(x)$ and $b^{s}(x)$ are given by:
\begin{equation}
a^{s}(x)=\frac{\delta l_{I}[\varphi]}{\delta\varphi(x)}|_{\varphi^{s}}
\label{as}
\end{equation}
\begin{equation}
b^{s}(x,x')=\frac{1}{2}\frac{\delta^{2} l_{I}[\varphi]}{\delta\varphi(x)\delta\varphi(x')}|_{\varphi^{s}}
\label{bs}
\end{equation}

\vspace{.3cm}

Substituting Eqs. (\ref{S01}) and (\ref{SI1}) into eq. (\ref{S}),
the full action $S[\varphi]$ can be written as
\begin{equation}
S[\varphi]=S[\varphi^{s}]+\delta S[\varphi^{s},\delta\varphi]
\label{S1}
\end{equation}
with
\begin{equation}
\nonumber \delta S[\varphi^{s},\delta\varphi]=S_{0}[\delta\varphi]+\int dx\,a^{s}(x)\delta\varphi(x)+\int dxdx'\,\delta\varphi(x)b^{s}(x,x')\delta\varphi(x')
\end{equation}
\begin{equation}
\delta S[\varphi^{s},\delta\varphi]=\int dxdx'\,\delta\varphi(x)[\delta(x-x')\,\frac{1}{2}\nabla_{x'}^{2}+b^{s}(x,x')]\delta\varphi(x')+\int dx\,a^{s}(x)\delta\varphi(x)
\label{deltaS}
\end{equation}
and where, in writing the previous equation, we have applied Eq.
(\ref{S0}).

We see from eq. (\ref{S1}) that the full action $S[\varphi]$ splits
into two contributions: the action $S[\varphi^{s}]$ for the slow
modes and a residual piece $\delta S[\varphi^{s},\delta\varphi]$
that mixes slow and fast modes. Performing a simple field
transformation that eliminates the linear term in
$\delta\varphi(x)$, the residual action is (at this order in the
perturbation theory around the slow modes) a quadratic theory for
the fast modes with a mass-like term given by $b^{s}(x,x')$ that
encodes the influence of the slow modes as well as that of the
interactions.

In order to derive an effective theory for the slow modes, we will
average the residual action $\delta S[\varphi^{s},\delta\varphi]$
with respect to the unperturbed ground state of the fast modes
$\delta\varphi$-field operator such that $\delta
S[\varphi^{s},\delta\varphi]$ will become a $\delta
S_{eff}[\varphi^{s}]$ and $S[\varphi]$ will become a
$S_{eff}[\varphi^{s}]$. As already pointed out, the R.G. procedure
corresponds to re-obtaining $S[\varphi]$ from $S_{eff}[\varphi^{s}]$
via establishing a ``renormalized" theory $S_{R}[\varphi]$ through a
scale renormalization of the theory's cutoff:
$\Lambda/s\rightarrow\Lambda$.

The average of the residual action will be more easily evaluated if
we express $\delta S[\varphi^{s},\delta\varphi]$ in terms of Green's
functions.

\subsection{General procedure II - Expressing $\delta S$ in terms of Green's functions}

The Green's function $G_{0}(x,x')$ for the free residual action is
defined through the equations:
\begin{eqnarray}
\left\{
\begin{array}{ll}
G_{0}^{-1}(x,x')=\delta(x-x')\,\frac{1}{2}\nabla_{x'}^{2}\\

\int dx''\,G_{0}^{-1}(x,x'')G_{0}(x'',x')=\delta(x-x')\\
\end{array}\right.
\label{G0xxprime}
\end{eqnarray}

\vspace{.3cm}

It follows from the definition that:
\begin{equation}
\nonumber \frac{1}{2}\nabla_{x}^{2}G_{0}(x,x')=\delta(x-x')
\end{equation}

From the above equation we see that $G_{0}(x,x')=G_{0}(x-x')$ and
thus we can Fourier transform the equation to write:
\begin{equation}
\nonumber \frac{1}{2}\nabla_{x}^{2}\int\frac{dq}{(2\pi)^{2}}\,G_{0}(q)e^{iq(x-x')}=\int\frac{dq}{(2\pi)^{2}}\,e^{iq(x-x')}
\end{equation}
\begin{equation}
\nonumber \int\frac{dq}{(2\pi)^{2}}\,G_{0}(q)\frac{1}{2}\left(-q^{2}-\frac{\omega^{2}}{v^{2}}\right)e^{iq(x-x')}=\int\frac{dq}{(2\pi)^{2}}\,e^{iq(x-x')}
\end{equation}
\begin{equation}
G_{0}(q)=G_{0}(q,\omega)=-\frac{2}{q^{2}+\omega^{2}/v^{2}}
\label{G0q}
\end{equation}

\vspace{.3cm}

The Green's function $G(x,x')$ for the full residual action is
defined through the equations
\begin{eqnarray}
\left\{
\begin{array}{ll}
G^{-1}(x,x')=G_{0}^{-1}(x,x')-\Sigma(x,x')\\

\int dx''\,G^{-1}(x,x'')G(x'',x')=\frac{1}{(2\pi)^{2}}\delta(x-x')\\
\end{array}\right.
\label{Gxxprime}
\end{eqnarray}

\vspace{.3cm}

where $\Sigma(x,x')$ is the theory's self-energy that accounts for
the corrections to the free Green's function due to interactions and
external fields.

Substituting the definition (\ref{G0xxprime}), it follows that:
\begin{equation}
\nonumber \frac{1}{2}\nabla_{x}^{2}G(x,x')-\int
dx''\,\Sigma(x,x'')G(x'',x')=\frac{1}{(2\pi)^{2}}\delta(x-x')
\end{equation}

\newpage

\begin{equation}
\nonumber
\int\frac{dq}{(2\pi)^{2}}\frac{dq'}{(2\pi)^{2}}\,G(q,q')\frac{1}{2}\left(-q^{2}-\frac{\omega^{2}}{v^{2}}\right)e^{iqx+iq'x'}\,-
\end{equation}

\begin{equation}
\nonumber -\int
dx''\int\frac{dq}{(2\pi)^{2}}\frac{dk}{(2\pi)^{2}}\frac{dk'}{(2\pi)^{2}}\frac{dq'}{(2\pi)^{2}}\,\Sigma(q,k)G(k',q')e^{iqx+i(k+k')x''+iq'x'}=
\end{equation}

\begin{equation}
\nonumber =\int \frac{dq}{(2\pi)^{4}}\,e^{iq(x-x')}
\end{equation}

\vspace{0.6cm}

\begin{equation}
\nonumber
\int\frac{dq}{(2\pi)^{2}}\frac{dq'}{(2\pi)^{2}}\,\left[\,G(q,q')G_{0}^{-1}(q)-\int\frac{dq''}{(2\pi)^{2}}\,\Sigma(q,q'')G(-q'',q')\,\right]e^{iqx+iq'x'}=
\end{equation}

\begin{equation}
\nonumber
=\int\frac{dq}{(2\pi)^{2}}\frac{dq'}{(2\pi)^{2}}\,\delta(q+q')e^{iqx+iq'x'}
\end{equation}

\vspace{.6cm}

\begin{equation}
G(q,q')=G_{0}(q)\delta(q+q')+G_{0}(q)\int\frac{dq''}{(2\pi)^{2}}\,\Sigma(q,q'')G(-q'',q')
\label{Gqqprime}
\end{equation}

\vspace{0.3cm}

The previous is the Dyson equation written in terms of the theory's
full self-energy. In our perturbation theory around the slow modes,
developed up to second order in the fast modes (which is analogous
to second order in a saddle-point expansion), the self energy is
simply given by:
\begin{equation}
\Sigma(q,q')=-b^{s}(q,q'),\quad \Sigma(x,x')=-b^{s}(x,x')
\label{Sigmaqqprime}
\end{equation}

Thus, Eq. (\ref{Gqqprime}) can be rewritten as:
\begin{equation}
G(q,q')=G_{0}(q)\delta(q+q')-G_{0}(q)\int\frac{dq''}{(2\pi)^{2}}\,b^{s}(q,q'')G(-q'',q')
\label{Gqqprime1}
\end{equation}

Now we can use Eq. (\ref{Gqqprime1}) to develop an expansion of
$G(q,q')$ in powers of the interaction coupling constant $g$.
\begin{equation}
\nonumber ..........
\end{equation}

$\star$ \emph{Perturbative expansion of $G(q,q')$ in powers of $g$}

At zero-th order in $g$ we have:
\begin{equation}
\nonumber G^{(0)}(q,q')=G_{0}(q)\delta(q+q')
\end{equation}

Up to first and second orders in $g$ we have, respectively:
\begin{equation}
\nonumber G^{(1)}(q,q')=G_{0}(q)\delta(q+q')-G_{0}(q)\int\frac{dq''}{(2\pi)^{2}}\,b^{s}(q,q'')G_{0}(-q'')\delta(-q''+q')
\end{equation}
\begin{equation}
\nonumber
G^{(1)}(q,q')=G_{0}(q)\delta(q+q')-\frac{1}{(2\pi)^2}G_{0}(q)b^{s}(q,q')G_{0}(-q')
\end{equation}

\begin{equation}
\nonumber
G^{(2)}(q,q')=G_{0}(q)\delta(q+q')-G_{0}(q)\int\frac{dq''}{(2\pi)^{2}}\,b^{s}(q,q'')[\,G_{0}(-q'')\delta(-q''+q')-
\end{equation}
\begin{equation}
\nonumber
\!\!\!\!\!\!\!\!\!\!\!\!\!\!\!\!\!\!\!\!\!\!\!\!\!\!\!\!\!\!\!\!\!\!\!\!-\frac{1}{(2\pi)^{2}}G_{0}(-q'')b^{s}(-q'',q')G_{0}(-q')\,]
\end{equation}

\begin{equation}
\nonumber
G^{(2)}(q,q')=G_{0}(q)\delta(q+q')-\frac{1}{(2\pi)^2}G_{0}(q)b^{s}(q,q')G_{0}(-q')+
\end{equation}
\begin{equation}
\nonumber\qquad\qquad\qquad\qquad\quad
+\frac{1}{(2\pi)^2}G_{0}(q)\int\frac{dq''}{(2\pi)^{2}}\,b^{s}(q,q'')G_{0}(-q'')b^{s}(-q'',q')G_{0}(-q')
\end{equation}

And so on...
\begin{equation}
\nonumber ..........
\end{equation}

Coming back to the residual action $\delta
S[\varphi^{s},\delta\varphi]$, applying the definitions
(\ref{G0xxprime}) and ((\ref{Gxxprime}) + Eq. (\ref{Sigmaqqprime}))
into Eq. (\ref{deltaS}), it follows that:
\begin{equation}
\nonumber \delta S[\varphi^{s},\delta\varphi]=\int dxdx'\,\delta\varphi(x)G^{-1}(x,x')\delta\varphi(x')+\int dx\,a^{s}(x)\delta\varphi(x)
\end{equation}

Finally, we can achieve a quadratic expression in the fast modes
$\delta\varphi$-field through a simple field transformation. Let:
\begin{equation}
\delta\varphi(x)=\bar{\varphi}(x)+r(x)
\label{fieldtransf}
\end{equation}

Then:
\begin{equation}
\nonumber \!\!\!\!\!\!\!\!\!\!\!\!\!\!\!\!\!\!\!\delta S[\varphi^{s},\delta\varphi]=\delta S[\varphi^{s},\bar{\varphi},r]=\int dxdx'\,\{\,\bar{\varphi}(x)G^{-1}(x,x')\bar{\varphi}(x')\,+
\end{equation}
\begin{equation}
\nonumber \qquad\qquad\qquad\qquad\qquad\qquad\qquad\qquad\qquad+\,\bar{\varphi}(x)[\,2G^{-1}(x,x')r(x')+a^{s}(x')\delta(x-x')\,]\,+
\end{equation}
\begin{equation}
\nonumber \qquad\qquad\qquad\qquad\qquad\qquad\qquad\qquad\qquad+\,r(x)[\,G^{-1}(x,x')r(x')+a^{s}(x')\delta(x-x')\,]\,\}
\end{equation}

Now if
\begin{equation}
\nonumber \int dx'\,G^{-1}(x,x')r(x')=-\frac{1}{2}a^{s}(x)
\end{equation}
i.e.,
\begin{equation}
r(x)=-\frac{1}{2}\int dx'\,G(x,x')a^{s}(x')
\label{r}
\end{equation}
then:
\begin{equation}
\delta S[\varphi^{s},\bar{\varphi}]=\int
dxdx'\,[\,\bar{\varphi}(x)G^{-1}(x,x')\bar{\varphi}(x')-\frac{1}{4}a^{s}(x)G(x,x')a^{s}(x')\,]
\label{deltaS1}
\end{equation}
%\begin{equation}
%\nonumber \Downarrow eqs.\,(\ref{Gxxprime}),\,(\ref{G0xxprime}),\,(\ref{S0})
%\end{equation}
%\begin{equation}
%\delta S[\varphi^{s},\bar{\varphi}]=S_{0}[\bar{\varphi}]+\int dxdx'\,\{\,\bar{\varphi}(x)b^{s}(x,x')\bar{\varphi}(x')-\frac{1}{4}a^{s}(x)G(x,x')a^{s}(x')\,\}
%\label{deltaS1}
%\end{equation}

\vspace{.3cm}

We are now ready to average $\delta S[\varphi^{s},\bar{\varphi}]$ in the fast modes' ground state.

\subsection{General procedure III - Averaging on the fast modes' ground state}

First, let us set up the preliminaries. Note that eqs. (\ref{r}),
(\ref{Gxxprime}) and (\ref{Sigmaqqprime}) imply that
$r(x)=r[\varphi^{s}(x)]$. Since, from eqs. (\ref{phi}) and
(\ref{fieldtransf}),
\begin{equation}
\nonumber \varphi(x)=\varphi^{s}(x)+r(x)+\bar{\varphi}(x)
\end{equation}
we can redefine the slow modes to incorporate the field $r(x)$ through a transformation $\varphi^{s}(x)+r(x)\rightarrow\varphi^{s}(x)$ and write:
\begin{equation}
\varphi(x)=\varphi^{s}(x)+\bar{\varphi}(x)
\label{phi1}
\end{equation}

Now, due to the same argument which led to eq. (\ref{S01}),
\begin{equation}
\nonumber H_{0}[\Pi,\varphi]=\int dx\,\frac{v}{2}(\Pi^{2}+(\partial_{x}\varphi)^{2})
\end{equation}
splits like:
\begin{equation}
H_{0}[\Pi,\varphi]=H_{0}[\Pi^{s},\varphi^{s}]+H_{0}[\bar{\Pi},\bar{\varphi}]
\label{H0}
\end{equation}

Let $|0\rangle^{\varphi}$, $|0\rangle^{\varphi^{s}}$ and $|0\rangle^{\bar{\varphi}}$ be, respectively, the full, slow modes and fast modes unperturbed ($H_{0}$'s) ground states. From eq. (\ref{H0}), we have:
\begin{equation}
|0\rangle^{\varphi}=|0\rangle^{\varphi^{s}}|0\rangle^{\bar{\varphi}}
\label{groundstates}
\end{equation}

The goal now is to rewrite the full unperturbed Green's function
\begin{equation}
\nonumber
G_{0}(x,x')\equiv\,^{\varphi}\langle0|\varphi(x)\varphi(x')|0\rangle^{\varphi}
\end{equation}
in the fast modes' subspace. So, using Eq. (\ref{phi1}), we can
write:
\begin{equation}
\nonumber
G_{0}(x,x')=\,^{\varphi}\langle0|\varphi^{s}(x)\varphi^{s}(x')|0\rangle^{\varphi}+\,^{\varphi}\langle0|\bar{\varphi}(x)\bar{\varphi}(x')|0\rangle^{\varphi}+
\end{equation}
\begin{equation}
\nonumber
\qquad\,\,\,\,\quad+\,^{\varphi}\langle0|\varphi^{s}(x)\bar{\varphi}(x')|0\rangle^{\varphi}+\,^{\varphi}\langle0|\bar{\varphi}(x)\varphi^{s}(x')|0\rangle^{\varphi}
\end{equation}

The last two terms in the previous equation vanish since the field
operators inside the brackets act on different subspaces. Scale
invariance implies that the first two terms must be equal, that is,
the space-time correlations do not depend on the field's
momentum-frequency scale. Therefore
\begin{equation}
\nonumber
G_{0}(x,x')=2\,^{\varphi}\langle0|\bar{\varphi}(x)\bar{\varphi}(x')|0\rangle^{\varphi}
\end{equation}
\begin{equation}
\nonumber
G_{0}(x,x')=2\,^{\bar{\varphi}}\langle0|\,^{\varphi^{s}}\langle0|\bar{\varphi}(x)\bar{\varphi}(x')|0\rangle^{\varphi^{s}}|0\rangle^{\bar{\varphi}}
\end{equation}
\begin{equation}
G_{0}(x,x')=2\,^{\bar{\varphi}}\langle0|\bar{\varphi}(x)\bar{\varphi}(x')|0\rangle^{\bar{\varphi}}
\label{G0xxprime1}
\end{equation}
where, in deriving of the previous equation, we have substituted Eq.
(\ref{groundstates}) and then used the fact that
$^{\varphi^{s}}\langle0|0\rangle\,^{\varphi^{s}}=1$.

The Fourier transforms of the unperturbed Green's function and of
the delta-functions are properly redefined in the fast modes'
subspace as
\begin{equation}
G_{0}(x,x')=G_{0}(x-x')=\int_{shell}\frac{dq}{(2\pi)^{2}}\,G_{0}(q)e^{iq(x-x')}
\label{newFD}
\end{equation}
\begin{equation}
\nonumber \delta(x)=\int_{shell}\frac{dq}{(2\pi)^{2}}\,e^{iqx}
\end{equation}
\begin{equation}
\delta(|q|-\Lambda/s)=\int dx\,e^{-iqx} \label{newDeltaF}
\end{equation}
i.e., constrained to the high momentum-frequency shell.

The effective slow modes' residual contribution to the action, let
us call it $\delta S_{eff}[\varphi^{s}]$, is obtained as the fast
modes' average of the residual $\delta
S[\varphi^{s},\bar{\varphi}]$:
\begin{equation}
\delta S_{eff}[\varphi^{s}]\equiv\,^{\bar{\varphi}}\langle0|\delta
S[\varphi^{s},\bar{\varphi}]|0\rangle^{\bar{\varphi}}
\label{deltaSeff}
\end{equation}

Substituting Eqs. (\ref{deltaS1}) and (\ref{G0xxprime1}) into Eq.
(\ref{deltaSeff}) we arrive at:
\begin{equation}
\nonumber \delta S_{eff}[\varphi^{s}]=\int
dxdx'\,\left[\,\frac{1}{2}G_{0}(x,x')G^{-1}(x,x')-\frac{1}{4}a^{s}(x)G(x,x')a^{s}(x')\,\right]
\end{equation}

Now, from Eqs. (\ref{S0}), (\ref{G0xxprime}), (\ref{Gxxprime}),
(\ref{Sigmaqqprime}) and (\ref{G0xxprime1}), it follows that:
\begin{equation}
\nonumber \delta S_{eff}[\varphi^{s}]=\int
dxdx'\,\left\{\,\frac{1}{2}G_{0}(x,x')[G_{0}^{-1}(x,x')+b^{s}(x,x')]-\frac{1}{4}a^{s}(x)G(x,x')a^{s}(x')\,\right\}
\end{equation}
\begin{equation}
\nonumber \delta
S_{eff}[\varphi^{s}]=\,^{\bar{\varphi}}\langle0|S_{0}[\tilde{\varphi}]|0\rangle^{\bar{\varphi}}+\int
dxdx'\,\left[\,\frac{1}{2}G_{0}(x,x')b^{s}(x,x')-\frac{1}{4}a^{s}(x)G(x,x')a^{s}(x')\,\right]
\end{equation}

The first term on the right hand side is just a constant and can be
absorbed through a trivial redefinition of $\delta
S_{eff}[\varphi^{s}]$, which can be finally written as
\begin{equation}
\delta S_{eff}[\varphi^{s}]=\int
dxdx'\,\left[\,\frac{1}{2}G_{0}(x,x')b^{s}(x,x')-\frac{1}{4}a^{s}(x)G_{0}(x,x')a^{s}(x')\,\right]
\label{deltaSeff1}
\end{equation}
where, in the last contribution to the integrand, $G(x,x')$ has been
replaced by $G_{0}(x,x')$ to keep terms only up to second order in
the $g$-coupling.

If we now replace $\delta S[\varphi^{s},\delta\varphi]$ in Eq.
(\ref{S1}) by $\delta S_{eff}[\varphi^{s}]$ given in Eq.
(\ref{deltaSeff1}), we write down the full slow modes' effective
action as:
\begin{equation}
S_{eff}[\varphi^{s}]=S[\varphi^{s}]+\delta S_{eff}[\varphi^{s}]
\label{Seff0}
\end{equation}

In order to compute the contribution of $\delta
S_{eff}[\varphi^{s}]$ to the effective theory, we need to explicit
the dependence of the integrand in Eq. (\ref{deltaSeff1}) on the
$\varphi^{s}$-field which is specific of each particular quantum
field theory. The following section performs this task for the
sine-Gordon model. The ultimate goal is to derive the model's
re-scaled action.

\subsection{Application I - The sine-Gordon model re-scaled action}

Let us compute the first term on the right hand side of Eq.
(\ref{deltaSeff1}):
\begin{equation}
F_{1}[\varphi^{s}]\equiv\int
dxdx'\,\frac{1}{2}G_{0}(x,x')b^{s}(x,x') \label{F1}
\end{equation}

Recalling the redefinition of the theory's Green function in the
fast modes' subspace Eq. (\ref{newFD}), it follows that:
\begin{equation}
\nonumber F_{1}[\varphi^{s}]=\frac{1}{2}\int
dxdx'\int_{shell}\frac{dq}{(2\pi)^{2}}\int\frac{dq'}{(2\pi)^{2}}\frac{dq''}{(2\pi)^{2}}\,G_{0}(q)b^{s}(q',q'')e^{iq(x-x')+iq'x+iq''x'}=
\end{equation}

\begin{equation}
\nonumber
F_{1}[\varphi^{s}]=\frac{1}{2}\int_{shell}\frac{dq}{(2\pi)^{2}}G_{0}(q)b^{s}(-q,q)
\end{equation}

\vspace{.6cm}

From Eqs. (\ref{bs}) and (\ref{SI}):
\begin{equation}
\nonumber
b^{s}(x,x')=-\frac{\beta^{2}}{2}g\cos(\beta\varphi^{s})\delta(x-x')=-\frac{\beta^{2}}{2}l_{I}[\varphi^{s}]\delta(x-x')
\end{equation}
\begin{equation}
\nonumber b^{s}(q,q')=\int dxdx'\,b^{s}(x,x')e^{iqx+iq'x'}
\end{equation}
\begin{equation}
b^{s}(q,q')=\int dx\,\left(-\frac{\beta^{2}}{2}l_{I}[\varphi^{s}]\right)e^{i(q+q')x}=b^{s}(q+q')
\label{bsqq'}
\end{equation}
\begin{equation}
\nonumber \Downarrow
\end{equation}
\begin{equation}
\nonumber
F_{1}[\varphi^{s}]=\frac{1}{2}b^{s}(q=0)\int_{shell}\frac{dq}{(2\pi)^{2}}\,G_{0}(q)
\end{equation}

Applying Eq. (\ref{G0q}),
\begin{equation}
\nonumber
F_{1}[\varphi^{s}]=-b^{s}(q=0)\int_{shell}\frac{dq}{(2\pi)^{2}}\,\frac{1}{q^{2}+\omega^{2}/v^{2}}
\end{equation}
\begin{equation}
\nonumber
F_{1}[\varphi^{s}]=-b^{s}(q=0)\int_{0}^{2\pi}\int_{\Lambda/s}^{\Lambda}\frac{d\theta
d|q|}{(2\pi)^{2}}\,\frac{1}{|q|}
\end{equation}
\begin{equation}
\nonumber F_{1}[\varphi^{s}]=-\frac{1}{2\pi}b^{s}(q=0)\ln(s)
\end{equation}
\begin{equation}
\nonumber F_{1}[\varphi^{s}]=\ln(s)\frac{\beta^{2}}{4\pi}\int
dx\,l_{I}[\varphi^{s}]
\end{equation}
where in the last step we have used Eq. (\ref{bsqq'}).

Finally, using Eq. (\ref{SI}), we get
\begin{equation}
F_{1}[\varphi^{s}]=-g\left(\frac{dl\beta^{2}}{4\pi}\right)\int
dx\,\cos(\beta\varphi^{s}) \label{resultF1}
\end{equation}
where:
\begin{equation}
dl\equiv-\ln(s) \label{dl}
\end{equation}

In order to computer the second term in the right hand side of Eq.
(\ref{deltaSeff1}), we expand $a^{s}(x')$ around $x'=x$ to write
\begin{equation}
F_{2}[\varphi^{s}]\equiv-\int dxdx'\,\frac{1}{4}G_{0}(x,x')a^{s}(x)a^{s}(x')
\label{F2}
\end{equation}
up to second order in the expansion, as:

\newpage

\begin{equation}
\nonumber \!\!\!\!\!\!\!\!\!\!\!\!\!\!\!\!\!\!\!\!\!\!\!\!\!\!\!\!\!\!\!\!\!\!\!\!\!\!\!\!\!\!\!\!\!\!\!\!\!\!\!\!\!\!\!\!\!\!\!\!\!\!\!\!\!\!\!\!\!\!\!\!\!\!\!\!\!\!F_{2}[\varphi^{s}]=-\frac{v^{2}}{4}\int dxdx'dtdt'\,G_{0}(x-x',t-t')\,\times
\end{equation}
\begin{equation}
\nonumber \!\!\!\!\!\!\!\!\!\!\!\!\!\!\!\!\!\!\!\!\!\!\!\!\times\,[\,(a^{s}(x,t))^{2}\,+
\end{equation}
\begin{equation}
\nonumber \qquad\qquad\qquad\qquad\qquad\qquad+\,a^{s}(x,t)\partial_{x}a^{s}(x,t)(x'-x)\,+\,a^{s}(x,t)\partial_{t}a^{s}(x,t)(t'-t)\,+
\end{equation}
\begin{equation}
\nonumber \qquad\qquad\qquad\qquad\qquad\qquad+\,\frac{1}{2}a^{s}(x,t)\partial_{x}^{2}a^{s}(x,t)(x'-x)^{2}\,+\,\frac{1}{2}a^{s}(x,t)\partial_{t}^{2}a^{s}(x,t)(t'-t)^{2}\,+
\end{equation}
\begin{equation}
\nonumber \qquad\qquad\qquad\,\,\,+\,a^{s}(x,t)\partial_{x}\partial_{t}a^{s}(x,t)(x'-x)(t'-t)\,]
\end{equation}

Or in a simpler form as:
\begin{equation}
\nonumber !\!\!\!\!\!\!\!\!\!\!\!\!\!\!\!\!\!\!\!\!\!\!\!\!\!\!\!\!\!\!\!\!\!\!\!\!\!\!\!\!\!\!\!\!\!\!\!\!\!\!\!\!\!\!\!\!\!\!\!\!F_{2}[\varphi^{s}]=-\frac{v^{2}}{4}\int dxdXdtdT\,G_{0}(X,T)\,\times
\end{equation}
\begin{equation}
\nonumber \qquad\quad\times\,[\,(a^{s}(x,t))^{2}\,-
\end{equation}
\begin{equation}
\nonumber \qquad\qquad\qquad\qquad\qquad\qquad\qquad\quad\,\,\,-\,Xa^{s}(x,t)\partial_{x}a^{s}(x,t)\,-\,Ta^{s}(x,t)\partial_{t}a^{s}(x,t)\,+
\end{equation}
\begin{equation}
\nonumber \,\,\,\,\,\,\,\,\,\qquad\qquad\quad\qquad\qquad\qquad\qquad\quad+\,\frac{1}{2}X^{2}a^{s}(x,t)\partial_{x}^{2}a^{s}(x,t)\,+\,\frac{1}{2}T^{2}a^{s}(x,t)\partial_{t}^{2}a^{s}(x,t)\,+
\end{equation}
\begin{equation}
\nonumber \qquad\qquad\qquad\qquad\quad+\,XTa^{s}(x,t)\partial_{x}\partial_{t}a^{s}(x,t)\,]
\end{equation}

Having Eqs. (\ref{newFD}) and (\ref{newDeltaF}) in mind, we have
that:
\begin{equation}
\nonumber G_{0}(q,\omega)=v\int dXdT\,G_{0}(X,T)e^{-i(qX+\omega T)}
\end{equation}
\begin{equation}
\nonumber \partial_{q}^{n}\partial_{\omega}^{m}G_{0}(q,\omega)=v\int dXdT\,(-i)^{n+m}X^{n}T^{m}G_{0}(X,T)e^{-i(qX+\omega T)}
\end{equation}
\begin{equation}
\nonumber
X^{n}T^{m}G_{0}(X,T)=\frac{1}{v}\int_{shell}\frac{dqd\omega}{(2\pi)^{2}}\,i^{n+m}\,\partial_{q}^{n}\partial_{\omega}^{m}G_{0}(q,\omega)e^{i(qX+\omega
T)}
\end{equation}
\begin{equation}
\!\!\!\!\!\!\!\!\!\!\!\!\!\!v\int
dXdT\,X^{n}T^{m}G_{0}(X,T)=\frac{i^{n+m}}{v}\int_{shell}
\frac{dqd\omega}{(2\pi)^{2}}\,\partial_{q}^{n}\partial_{\omega}^{m}G_{0}(q,\omega)\delta(|q|-\Lambda/s)\equiv\,f_{\partial_{q}^{n},\partial_{\omega}^{m}}(\Lambda/s)
\label{f}
\end{equation}

\newpage

Therefore:
\begin{equation}
\nonumber
\!\!\!\!\!\!\!\!\!\!\!\!\!\!\!\!\!\!\!\!\!\!\!\!\!\!\!\!\!\!\!\!\!\!\!\!\!\!\!\!\!\!\!\!\!\!\!\!\!\!\!\!\!\!\!\!\!\!\!\!\!\!\!\!\!\!\!\!\!\!\!\!\!\!\!\!\!\!\!\!\!\!\!\!\!\!\!\!\!\!\!\!\!\!\!\!\!\!\!\!F_{2}[\varphi^{s}]=-\frac{v}{4}\int
dxdt\,[\,f_{\partial_{q}^{0}\partial_{\omega}^{0}}(\Lambda/s)(a^{s}(x,t))^{2}\,-
\end{equation}
\begin{equation}
\nonumber \qquad\qquad\qquad\quad\!
-\,f_{\partial_{q}^{1}\partial_{\omega}^{0}}(\Lambda/s)\,a^{s}(x,t)\partial_{x}a^{s}(x,t)\,-\,f_{\partial_{q}^{0}\partial_{\omega}^{1}}(\Lambda/s)\,a^{s}(x,t)\partial_{t}a^{s}(x,t)\,+
\end{equation}
\begin{equation}
\nonumber \qquad\qquad\quad\,\,\,\,\,\,\,\,\,\,
+\,\frac{1}{2}f_{\partial_{q}^{2}\partial_{\omega}^{0}}(\Lambda/s)\,a^{s}(x,t)\partial_{x}^{2}a^{s}(x,t)\,+\,\frac{1}{2}f_{\partial_{q}^{0}\partial_{\omega}^{2}}(\Lambda/s)\,a^{s}(x,t)\partial_{t}^{2}a^{s}(x,t)\,+
\end{equation}
\begin{equation}
\nonumber
\!\!\!\!\!\!\!\!\!\!\!\!\!\!\!\!\!\!\!\!\!\!\!\!\!\!\!\!\!\!+\,f_{\partial_{q}^{1}\partial_{\omega}^{1}}(\Lambda/s)\,a^{s}(x,t)\partial_{x}\partial_{t}a^{s}(x,t)\,]
\end{equation}

\vspace{.3cm}

From Eqs. (\ref{as}) and (\ref{SI}), it follows that:
\begin{equation}
\nonumber a^{s}(x)=-\beta g\sin(\beta\varphi^{s})
\end{equation}
\begin{equation}
\nonumber (a^{s}(x))^{2}=(\beta g)^{2}\sin^{2}(\beta\varphi^{s})=\frac{(\beta g)^{2}}{2}[1-\cos(2\beta\varphi^{s})]
\end{equation}
\begin{equation}
\nonumber a^{s}(x)\partial_{x,t}a^{s}(x)=\beta(\beta g)^{2}\sin(\beta\varphi^{s})\cos(\beta\varphi^{s})\partial_{x,t}\varphi^{s}=\frac{\beta^{3} g^{2}}{2}\sin(2\beta\varphi^{s})\partial_{x,t}\varphi^{s}
\end{equation}
\begin{equation}
\nonumber a^{s}(x)\partial_{x,t}^{2}a^{s}(x)=\beta(\beta g)^{2}\sin(\beta\varphi^{s})[-\beta\sin(\beta\varphi^{s})(\partial_{x,t}\varphi^{s})^{2}+\cos(\beta\varphi^{s})\partial_{x,t}^{2}\varphi^{s}]
\end{equation}
\begin{equation}
\nonumber
\qquad\qquad\quad\,\,\,\,=-\frac{\beta^{4}g^{2}}{2}[1-\cos(2\beta\varphi^{s})](\partial_{x,t}\varphi^{s})^{2}+\frac{\beta^{3}g^{2}}{2}\sin(2\beta\varphi^{s})\partial_{x,t}^{2}\varphi^{s}
\end{equation}
\begin{equation}
\nonumber
a^{s}(x)\partial_{x}\partial_{t}a^{s}(x)=-\frac{\beta^{4}g^{2}}{2}[1-\cos(2\beta\varphi^{s})](\partial_{x}\varphi^{s})(\partial_{t}\varphi^{s})+\frac{\beta^{3}g^{2}}{2}\sin(2\beta\varphi^{s})\partial_{x}\partial_{t}\varphi^{s}
\end{equation}

\vspace{.3cm}

Keeping only the non-oscillatory contributions (since the
oscillatory ones average to zero when integrated in space and time),
we have
\begin{equation}
\nonumber F_{2}[\varphi^{s}]=-\frac{(\beta
g)^{2}V}{8}f_{\partial_{q}^{0}\partial_{\omega}^{0}}(\Lambda/s)\,+
\end{equation}
\begin{equation}
\nonumber \!\!\!\!\!+\,\frac{\beta^{4}g^{2}v}{8}\int
dxdt\,\left[\,\frac{1}{2}f_{\partial_{q}^{2}\partial_{\omega}^{0}}(\Lambda/s)(\partial_{x}\varphi^{s})^{2}+\frac{1}{2}f_{\partial_{q}^{0}\partial_{\omega}^{2}}(\Lambda/s)(\partial_{t}\varphi^{s})^{2}+f_{\partial_{q}^{1}\partial_{\omega}^{1}}(\Lambda/s)(\partial_{x}\varphi^{s})(\partial_{t}\varphi^{s})\,\right]
\end{equation}
where $V\equiv v\int\int dxdt$ is the system's volume in space and
time.

Now, from Eq. (\ref{f}),
\begin{equation}
\nonumber
f_{\partial_{q}^{2}\partial_{\omega}^{0}}(\Lambda/s)=-\frac{1}{v}\int_{shell}\frac{dqd\omega}{(2\pi)^{2}}\,\partial_{q}^{2}G_{0}(q,\omega)\delta(|q|-\Lambda/s)
\end{equation}
\begin{equation}
\nonumber
f_{\partial_{q}^{0}\partial_{\omega}^{2}}(\Lambda/s)=-\frac{1}{v}\int_{shell}\frac{dqd\omega}{(2\pi)^{2}}\,\partial_{\omega}^{2}G_{0}(q,\omega)\delta(|q|-\Lambda/s)
\end{equation}
\begin{equation}
\nonumber
f_{\partial_{q}^{1},\partial_{\omega}^{1}}(\Lambda/s)=-\frac{1}{v}\int_{shell}\frac{dqd\omega}{(2\pi)^{2}}\,\partial_{q}\partial_{\omega}G_{0}(q,\omega)\delta(|q|-\Lambda/s)
\end{equation}

\vspace{.3cm}

and, from Eq. (\ref{G0q}):
\begin{equation}
\nonumber \partial_{q}^{2}G_{0}(q,\omega)=\frac{4}{(q^{2}+\omega^{2}/v^{2})^{2}}\left(1-\frac{4q^{2}}{q^{2}+\omega^{2}/v^{2}}\right)
\end{equation}
\begin{equation}
\nonumber \partial_{\omega}^{2}G_{0}(q,\omega)=\frac{4}{v^{2}(q^{2}+\omega^{2}/v^{2})^{2}}\left(1-\frac{4\omega^{2}/v^{2}}{q^{2}+\omega^{2}/v^{2}}\right)
\end{equation}
\begin{equation}
\nonumber \partial_{q}\partial_{\omega}G_{0}(q,\omega)=-\frac{16q\omega/v^{2}}{(q^{2}+\omega^{2}/v^{2})^{3}}
\end{equation}

\vspace{.3cm}

Making a change to polar coordinates:
\begin{equation}
\nonumber
f_{\partial_{q}^{2}\partial_{\omega}^{0}}(\Lambda/s)=-\int_{0}^{2\pi}\int_{\Lambda/s}^{\Lambda}\frac{d\theta
d|q||q|}{(2\pi)^{2}}\,\frac{4}{|q|^{4}}(1-4\cos^{2}(\theta))\,\delta(|q|-\Lambda/s)=\frac{2s^{3}}{\pi\Lambda^{3}}
\end{equation}
\begin{equation}
\nonumber
f_{\partial_{q}^{0}\partial_{\omega}^{2}}(\Lambda/s)=-\int_{0}^{2\pi}\int_{\Lambda/s}^{\Lambda}\frac{d\theta
d|q||q|}{(2\pi)^{2}}\,\frac{4}{v^{2}|q|^{4}}(1-4\sin^{2}(\theta))\,\delta(|q|-\Lambda/s)=\frac{1}{v^{2}}\frac{2s^{3}}{\pi\Lambda^{3}}
\end{equation}
\begin{equation}
\nonumber
f_{\partial_{q}^{1}\partial_{\omega}^{1}}(\Lambda/s)=\int_{0}^{2\pi}\int_{\Lambda/s}^{\Lambda}\frac{d\theta
d|q||q|}{(2\pi)^{2}}\,\frac{16}{v|q|^{4}}\cos(\theta)\sin(\theta)\,\delta(|q|-\Lambda/s)=0
\end{equation}

\vspace{.3cm}

Absorbing the constant term in the previous computation of
$F_{2}[\varphi^{s}]$ and performing an integration by parts, we
have:
\begin{equation}
\nonumber
F_{2}[\varphi^{s}]=-\frac{\beta^{4}g^{2}s^{3}v}{8\pi\Lambda^{3}}\int
dxdt\,[\,\varphi^{s}\partial_{x}^{2}\varphi^{s}+\varphi^{s}\frac{1}{v^{2}}\partial_{t}^{2}\varphi^{s}\,]
\end{equation}

\begin{equation}
\nonumber
F_{2}[\varphi^{s}]=-\frac{\beta^{4}g^{2}s^{3}}{8\pi\Lambda^{3}}\int
dx\,\varphi^{s}\nabla_{x}^{2}\varphi^{s}
\end{equation}

\vspace{.3cm}

Using Eq. (\ref{dl}) and expanding around $dl=0$ ($s=1$), we obtain
\begin{equation}
\nonumber
F_{2}[\varphi^{s}]=-\left(\frac{\beta^{4}g^{2}}{8\pi\Lambda^{3}}-\frac{dl3\beta^{4}g^{2}}{8\pi\Lambda^{3}}\right)\int
dx\,\varphi^{s}\nabla_{x}^{2}\varphi^{s}
\end{equation}
\begin{equation}
F_{2}[\varphi^{s}]=\frac{dl3\beta^{4}g^{2}}{8\pi\Lambda^{3}}\int
dx\,\varphi^{s}\nabla_{x}^{2}\varphi^{s} \label{resultF2}
\end{equation}
where we have used the fact that the first term in the parenthesis
goes to zero in the limit of large momentum-frequency cutoff
$\Lambda$.

Substituting the results (\ref{resultF1}) and (\ref{resultF2}) for
the quantities (\ref{F1}) and (\ref{F2}) into eq. (\ref{deltaSeff1})
gives:
\begin{equation}
\nonumber \delta S_{eff}[\varphi^{s}]=\int
dx\,\left[\,\left(\frac{dl3\beta^{4}g^{2}}{8\pi\Lambda^{3}}\right)\varphi^{s}\nabla_{x}^{2}\varphi^{s}-g\left(\frac{dl\beta^{2}}{4\pi}\right)\cos(\beta\varphi^{s})\,\right]
\end{equation}

\vspace{.3cm}

Then, according to Eqs. (\ref{Seff0}) and (\ref{S})-(\ref{SI}), the
full slow modes' effective action can be finally written as:
\begin{equation}
S_{eff}[\varphi^{s}]=\int
dx\,\left[\,\frac{1}{2}\left(1+\frac{dl3\beta^{4}g^{2}}{4\pi\Lambda^{3}}\right)\varphi^{s}\nabla_{x}^{2}\varphi^{s}+g\left(1-\frac{dl\beta^{2}}{4\pi}\right)\cos(\beta\varphi^{s})\,\right]
\label{Seff}
\end{equation}

\vspace{.3cm}

To go back to original scale where $\varphi^{s}\rightarrow\varphi$ we perform the substitutions
\begin{equation}
\nonumber (x,vt)\rightarrow(\frac{x}{s},\frac{vt}{s})
\end{equation}
\begin{equation}
\nonumber \Downarrow
\end{equation}
\begin{equation}
\nonumber \int dx\rightarrow \frac{1}{s^{2}}\int dx; \quad
\nabla^{2}_{x}\rightarrow s^{2}\nabla^{2}_{x}
\end{equation}
in the above effective slow modes' action. Therefore, the re-scaled
action is:
\begin{equation}
S_{R}[\varphi]=\int
dx\,\left[\,\frac{1}{2}\left(1+\frac{dl3\beta^{4}g^{2}}{4\pi\Lambda^{3}}\right)\varphi\nabla_{x}^{2}\varphi+gs^{-2}\left(1-\frac{dl\beta^{2}}{4\pi}\right)\cos(\beta\varphi)\,\right]
\label{SR}
\end{equation}

\vspace{.3cm}

Under the field transformation $\beta\varphi\rightarrow\varphi$, the re-scaled and original actions become:
\begin{equation}
S_{R}[\varphi]=\int
dx\,\left[\,\frac{1}{2\beta^{2}}\left(1+\frac{dl3\beta^{4}g^{2}}{4\pi\Lambda^{3}}\right)\varphi\nabla_{x}^{2}\varphi+gs^{-2}\left(1-\frac{dl\beta^{2}}{4\pi}\right)\cos(\varphi)\,\right]
\label{SR2}
\end{equation}
\begin{equation}
S[\varphi]=\int
dx\,\left[\,\frac{1}{2\beta^{2}}\varphi\nabla_{x}^{2}\varphi+g\cos(\varphi)\,\right]
\label{S2}
\end{equation}

\subsection{Application II - The sine-Gordon model flow equations}

The R.G. statement regarding the theory's scale invariance amounts
to matching Eqs. (\ref{SR2}) and (\ref{S2}). As a result of this
equivalence one writes the model's renormalized parameters in terms
of the bare ones and of the re-scaling parameter in the following
way:
\begin{equation}
\nonumber\beta_{R}^{-2}=\beta^{-2}\left(1+\frac{dl3\beta^{4}g^{2}}{4\pi\Lambda^{3}}\right),\qquad
g_{R}=gs^{-2}\left(1-\frac{dl\beta^{2}}{4\pi}\right) \label{betaR}
\end{equation}

\vspace{.3cm}

Defining the differential of a parameter as $dX\equiv X_{R}-X$, we can rewrite the above equations in differential form as:
\begin{equation}
\nonumber d\beta^{-2}=\frac{3\beta^{2}g^{2}}{4\pi\Lambda^{3}}dl
\end{equation}
\begin{equation}
\nonumber \frac{d\beta^{-2}}{dl}=-(\beta^{2})^{-2}\frac{d\beta^{2}}{dl}=\frac{3\beta^{2}g^{2}}{4\pi\Lambda^{3}}
\end{equation}
\begin{equation}
\nonumber \frac{d\beta^{2}}{dl}=-\frac{3\beta^{6}g^{2}}{4\pi\Lambda^{3}}
\end{equation}

\begin{equation}
\nonumber g_{R}=g(1+2dl)\left(1-\frac{dl\beta^{2}}{4\pi}\right)
\end{equation}
\begin{equation}
\nonumber dg=g\left(2-\frac{\beta^{2}}{4\pi}\right)dl
\end{equation}
\begin{equation}
\nonumber \frac{dg}{dl}=2g\left(1-\frac{\beta^{2}}{8\pi}\right)
\end{equation}

The R.G. flow equations for the sine-Gordon model parameters are
given in terms of the scale $l=-\ln(s)+l_{0}$ by:
\begin{eqnarray}
\left\{
\begin{array}{ll}
\frac{du}{dl}=2u(1-K)\\

\frac{dK}{dl}=-u^{2}K^{3}\\
\end{array}\right.
\label{RGflows}
\end{eqnarray}
where:
\begin{equation}
K=\frac{\beta^{2}}{8\pi}\qquad u=4\sqrt{\frac{3\pi}{\Lambda^{3}}}g
\label{Kandu}
\end{equation}

\newpage

\section{Kosterlitz-Thouless Phase Diagram}

\subsection{Analysis of the flow equations}

A possible solution for the system of first order coupled
differential equations given by Eqs. (\ref{RGflows}) can be depicted
as a path $(K(l),u(l))$ in the $K-u$ plane, where $l$ is a
parametric running variable. Each such path describes the flow of
the initial point $(K_{0},u_{0})\equiv(K(l_{0}),u(l_{0}))$ when the
variable $l$ starts at $l_{0}$ and runs in the direction of the
model's original scale (i.e. to recover the full momentum-frequency
space). The phase diagram of Eqs. (\ref{RGflows}) is given by the
collection of all possible paths $(K(l),u(l))$ in the $K-u$ plane.

We start the analysis by observing that Eqs. (\ref{RGflows}) imply
$du/dl=dK/dl=0$ for $u=0$. We say that the phase diagram has a line
of fixed points at $u=0$ meaning that the flow stops when and if it
hits that line for some $l=l^{*}$. In this case, the system
parameters will take on the value $(K(l^{*}),0)$ for any scale
$l\geq l^{*}$. In particular, a system with bare parameters
$(K_{0},0)$ will not flow at all, i.e., a free quadratic model does
not get renormalized under a scale transformation, as expected.

Secondly, we see from the first Eq. (\ref{RGflows}) that the flow of
the coupling $u$: (i) points upward inside the $K<1$ half of the
phase diagram and downward inside the $K>1$ half if $u>0$; (ii)
points downward inside the $K<1$ half of the phase diagram and
upward inside the $K>1$ half if $u<0$; (iii) points horizontally at
$K=1$.

Therefore, the fixed points along the line $u=0$ are unstable for
$K<1$ (a point just above or below the segment $\{K<1,u=0\}$ will
flow away from it) and stable for $K>1$ (a point just above or below
the segment $\{K>1,u=0\}$ will flow towards it).

The second Eq. (\ref{RGflows}) implies that the flow of the
parameter $K$: (iv) points to the left inside the $K>0$ half of the
phase diagram and (v) points to the right inside the $K<0$ half,
regardless of the value of $u$, (vi) stops whenever $K$ reaches the
line $K=0$. In this case, according to items (i) and (ii), $u$ flows
up (vertically) for $u>0$ and down (vertically) for $u<0$. In
particular, this shows that no path can possibly cross the $K=0$
line where the flow of $K$ changes direction.

Combining the above conclusions, we see that there are three possible regions in the phase diagram:\\

1) \emph{Strong coupling regime}: The region of paths $(K(l),u(l))$
constrained to the $K<1$ half of the phase diagram and which flow to
the regime of large $|u|$ and small $|K|$. In this regime, the
interaction, whose strength is proportional to the value
of $|u|$, is said to be relevant.\\

2) \emph{Vanishing coupling regime}: The region of paths
$(K(l),u(l))$ constrained to the $K>1$ half of the phase diagram and
which flow to the regime of vanishing $u$ and fixed $K$. In this
regime, the interaction is irrelevant.\\

3) \emph{Crossover regime}: The region of paths $(K(l),u(l))$ that
go from the $K>1$ into the $K<1$ half of the phase diagram, thus
initially flowing towards a minimum value of $|u|$ attained at $K=1$
(where $du/dl=0$). Past this point these paths turn into the regime
of large $|u|$ and small positive $K$. In this case, the interaction
is said to be marginal.

Notice that, according to item (iv) above, there is no region for
paths going in the opposite direction, i.e. from $0<K<1$ to $K>1$.

Let us complement this discussion with a simple algebraic analysis.

We focus on the region around the line $K=1$ which is where the
interesting physics takes place. Then writing
\begin{equation}
K=1+v \label{K}
\end{equation}
we can rewrite Eqs. (\ref{RGflows}) up to first order in $v$ as:
\begin{eqnarray}
\nonumber \left\{
\begin{array}{ll}
\frac{du}{dl}=-2uv\\

\frac{dv}{dl}=-u^{2}(1+3v)\\
\end{array}\right.
\end{eqnarray}
\begin{eqnarray}
\nonumber \left\{
\begin{array}{ll}
u\frac{du}{dl}=-2u^{2}v\Rightarrow\frac{du^{2}}{dl}=-4u^{2}v\\

v\frac{dv}{dl}=-u^{2}v\Rightarrow\frac{dv^{2}}{dl}=-2u^{2}v\\
\end{array}\right.
\end{eqnarray}

\begin{equation}
\nonumber
\frac{d}{dl}(u^{2}-2v^{2})=\frac{d}{dl}(u^{2}-2(K-1)^{2})=0
\end{equation}

\vspace{.3cm}

The quantity $u^{2}-2(K-1)^{2}$ is an invariant for each solution
$(K(l),u(l))$, i.e.,
\begin{equation}
u^{2}(l)-2(K(l)-1)^{2}=c \label{invariant}
\end{equation}
where $c$ is a constant (for a given solution) that can be
determined, for example, by the initial conditions:
$c=u_{0}^{2}-2(K_{0}-1)^{2}$.

Now, let $(K_{f},|u|\rightarrow0)$ be the extreme point of a path
$(K(l),u(l))$ that flows to or from the line of fixed points $u=0$.
Eq. (\ref{invariant}) implies that:
\begin{equation}
\nonumber K_{f}=1\pm\sqrt{\frac{-c}{2}}
\end{equation}

We see that for $K_{f}$ to exist we must have $c\leq0$; otherwise
the path does not flow to or from the line of fixed points. To be
more precise, there are three possible cases:\\

$\star$ $c=0$
\begin{equation}
\nonumber u(l)=\pm\sqrt{2}(K(l)-1)
\end{equation}
\begin{equation}
\nonumber fixed\,point\,\rightarrow\,(K_{f},u)=(1,0)
\end{equation}

\vspace{.8cm}

$\star$ $c<0$
\begin{equation}
\nonumber |u(l)|<\sqrt{2}|K(l)-1|
\end{equation}
\begin{equation}
\nonumber
fixed\,points\,\rightarrow\,(K_{f},u)=(1\pm\sqrt{\frac{-c}{2}},0)
\end{equation}

\vspace{.8cm}

$\star$ $c>0$
\begin{equation}
\nonumber |u(l)|>\sqrt{2}|K(l)-1|
\end{equation}
\begin{equation}
\nonumber \nexists\,fixed\,points
\end{equation}

\vspace{.5cm}

Based on the previous qualitative and quatitative analysis, we can
draw the phase diagram for the sine-Gordon model as in Fig. 1. This
is the known Kosterlitz-Thouless (K-T) phase diagram. The straight
lines $u=\pm\sqrt{2}(K-1)$, given by the condition $c=0$, define the
boundaries between the strong coupling, the vanishing coupling and
the crossover regimes. These lines are called ``separatrices". The
strong and vanishing coupling regimes consist of the family of
hyperbolas defined by the condition $c<0$ (and thus ``enclosed" by
the separatrices) while the crossover regime corresponds to the
hyperbolas defined by $c>0$ (``outside" the separatrices).
\begin{figure}[h]
\begin{center}
\includegraphics[scale=1.5]{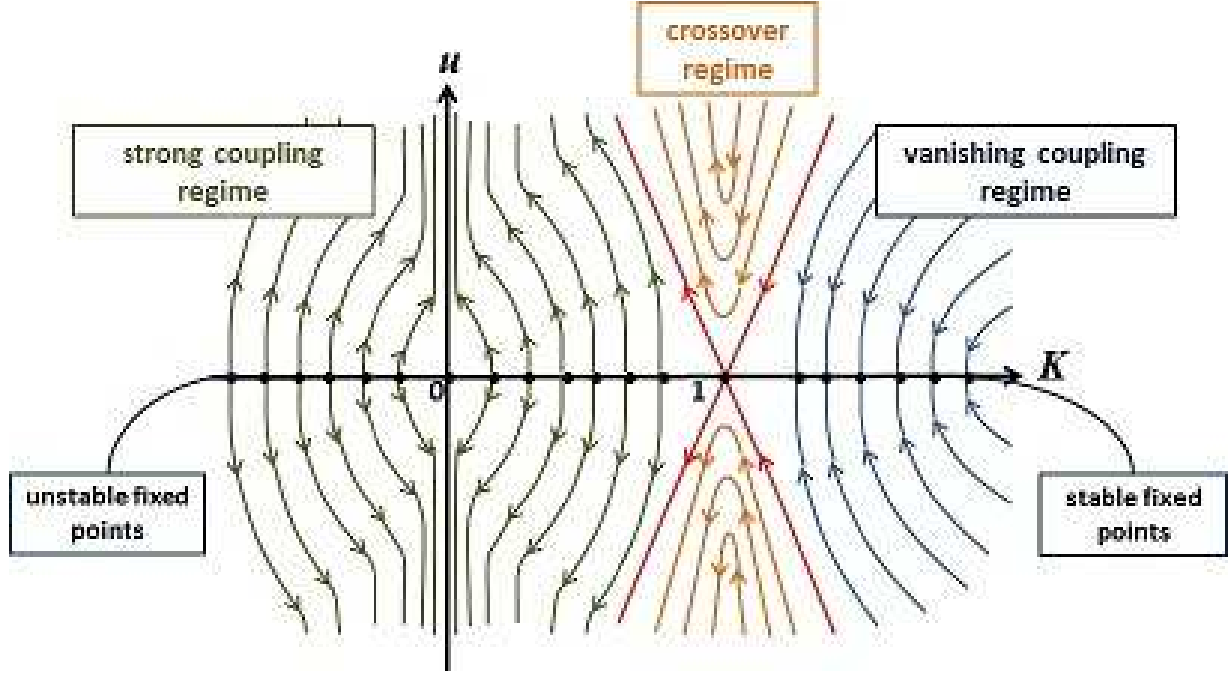}
\end{center}
\caption{Kosterlitz-Thouless phase diagram for the sine-Gordon
model.}
\end{figure}

\subsection{Gap}

In both the strong coupling and the crossover regimes the flows are
towards large $|u|$. At some critical scale in these flows, call it
$l_{c}$, the interaction becomes too strong, driving a phase
transition in the system. Thus, at $l_{c}$, the system loses scale
invariance and the R.G. statement is no longer valid. Based on the
perturbative nature of the R.G. procedure, a reasonable estimate for
$l_{c}$ is the scale at which the flow of $|u|$ reaches unity. The
system's critical correlation length $\xi_{c}$ can be assessed
through the expression: $\xi_{c}\propto\exp(l_{c})$.

Since the gap $\Delta\propto\xi_{c}^{-1}$,
\begin{equation}
\nonumber \Delta=\exp(-\,l_{c})
\end{equation}
gives an estimate for the gap (except for a multiplicative energy
factor) that opens up in a system that starts at $|u_{0}|<1$ and
flows to the large $|u|$-regime.

\newpage

Our task is to determine the value of $l_{c}$, and thus of $\Delta$,
as a function of the sine-Gordon bare parameters $K_{0}$ and
$|u_{0}|<1$. A first approximation is to consider the perturbative
R.G. only up to first order in the coupling constant $g$. At this
order, we can straightforwardly integrate the flow equation for $u$
and determine $l_{c}$.

Keeping corrections only up to first order in $g$ would have led to
simplified flow equations of the form:
\begin{eqnarray}
\nonumber \left\{
\begin{array}{ll}
\frac{du}{dl}=2u(1-K)\\

\frac{dK}{dl}=0\\
\end{array}\right.
\end{eqnarray}
as can be seen directly from Eqs. (\ref{RGflows}) and (\ref{Kandu})
by taking $u^{2}\rightarrow0$.

Since now $K=K_{0}$ is a constant parameter, we can write:
\begin{equation}
\nonumber
\int_{u_{0}}^{u}\frac{du'}{u'}=2(1-K_{0})\int_{l_{0}}^{l}dl'
\end{equation}
\begin{equation}
\nonumber u(l)=u_{0}\exp[2(1-K_{0})(l-l_{0})]
\end{equation}

We see that, for $K_{0}<1$, $|u|$ increases boundlessly, while for
$K_{0}>1$, $|u|$ decreases until it reaches the line of fixed points
$u=0$. Note that first order perturbative R.G. cannot capture
crossover paths. In particular, $K=1$ represents a line of fixed
points at this level of approximation. Just to illustrate, the
sine-Gordon model phase diagram produced by first order perturbative
R.G. looks as in Fig. 2.
\begin{figure}[h]
\begin{center}
\includegraphics[scale=1.2]{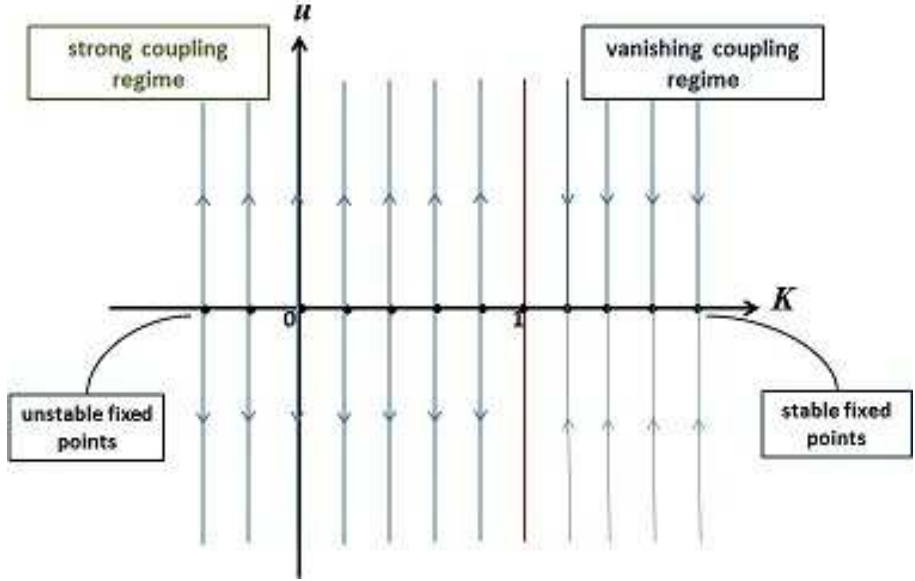}
\end{center}
\caption{Phase diagram for the sine-Gordon model produced by first
order perturbative R.G.}
\end{figure}

Coming back to the gap, for $K_{0}<1$ and $|u_{0}|<1$ we can write:
\begin{equation}
\nonumber 1=|u_{0}|\exp[2(1-K_{0})(l_{c}-l_{0})]
\end{equation}
\begin{equation}
\nonumber l_{c}=l_{0}+\ln(\,|u_{0}|^{1/2(K_{0}-1)}\,)
\end{equation}

Therefore, in first order approximation:
\begin{eqnarray}
\Delta=\Delta(K_{0},u_{0})=\left\{
\begin{array}{ll}
c_{0}|u_{0}|^{1/2(1-K_{0})}\qquad K_{0}\leq1\,(and\,|u_{0}|<1)\\

0\qquad\qquad\qquad\quad K_{0}\geq1\\
\end{array}\right.
\label{gap}
\end{eqnarray}

As shown in Fig. 3, for a given $|u_{0}|<1$, the gap decreases with
$K_{0}$ (since $|u_{0}|<1$) until it reaches zero at the critical
value $K_{0}^{c}=1$. On the other hand, given $K_{0}<1$, the gap
increases with $|u_{0}|$ until it reaches its maximum value of
$c_{0}=\exp(-l_{0})$ corresponding to $|u_{0}|=1$. This behavior of
the gap with $K_{0}$ and $|u_{0}|$ is an expression of the fact that
the critical correlation length decreases as one goes deeper into
the strong coupling regime, i.e. as $|u_{0}|$ increases and $K_{0}$
decreases.
\begin{figure}[h]
\begin{center}
\includegraphics[scale=1.4]{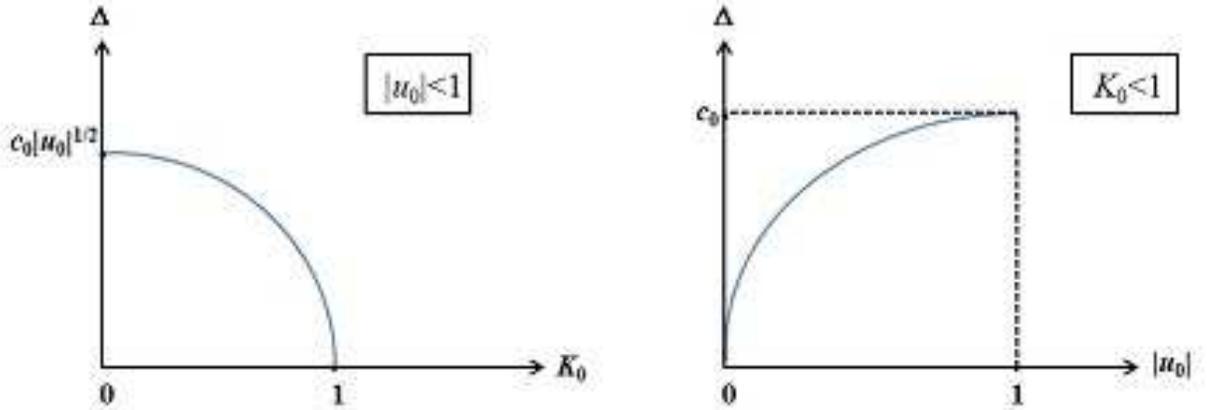}
\end{center}
\caption{First order gap as a function of $K_{0}$ (left) and
$|u_{0}|$ (right).}
\end{figure}

The line $K=1$ in Fig. 2 defines the boundary between the gapless
(vanishing coupling regime) and gapped (strong coupling regime)
regions of the phase diagram. The system can undergo a phase
transition between the gapless and gapped phases by varying the
parameter $K_{0}$ across the line $K=1$.

\newpage

A first correction to the first order gap of Eq. (\ref{gap}) and
Fig. 3 can be achieved by expanding the gapped region into the
crossover regime of Fig. 1 (where marginal paths may start at the
region $K>1$ but ultimately flow into the large $|u|$ regime). This
correction should take into account that the boarder line in the K-T
phase diagram is no longer given by $K=1$, but by $K=1\pm
u/\sqrt{2}$, where the upper sign stands for $u>0$ while the lower
one is for $u<0$.

Based on this and guided by the first-order results of Fig. 3, we
can draw a qualitative picture for the gap produced by second-order
perturbative R.G. such as depicted in Fig 4.
\begin{figure}[h]
\begin{center}
\includegraphics[scale=1.4]{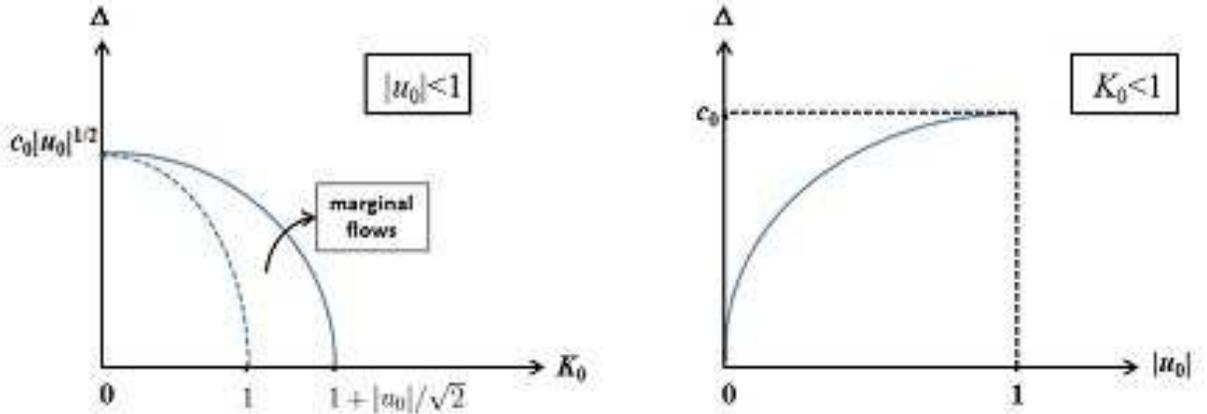}
\end{center}
\caption{Qualitative picture for the second-order gap as a function
of $K_{0}$ (left) and $|u_{0}|$ (right).}
\end{figure}

For a given $|u_{0}|<1$, the gap decreases with $K_{0}$ until it
reaches zero at the critical value $K_{0}^{c}=1+|u_{0}|/\sqrt{2}$.
The dashed line on the graph represents the gap as given by first
order R.G. The region between the two curves accounts for the
contribution of marginal paths to the gap opening. As indicated by
the second Eq. (\ref{RGflows}), for small $K_{0}$, the parameter $K$
remains almost constant along the second order R.G. flow. In other
words, close to the line $K=0$, the second order flow is essentially
vertical like in first order. Therefore, for small $K_{0}$, the
first- and second-order approximations should give roughly the same
results for the gap, as shown in Fig. 4. As $K_{0}$ increases, the
first and second-order gaps depart from each other to die in
different critical points.

Now, given $K_{0}<1$, the dependance of the gap on $|u_{0}|$ should
be similar to that of the first order approximation. It is not
possible to apply a similar qualitative reasoning for the $K_{0}>1$
region because flows starting there have a non-trivial marginal
behavior. In particular, since the marginal flows are ``longer", the
critical scale $l_{c}$ (at which a gap would open up) might exceed
the system's cutoff and, in practice, the phase transition might not
be realizable. The important point anyway is that, deep inside the
strong coupling sector of the K-T phase diagram where the gap is
more relevant (larger), $\Delta=c_{0}|u_{0}|^{1/2(1-K_{0})}$ remains
a good quantitative estimate.

\newpage

\section{Applications in Condensed Matter Physics}

The main motivation of the sine-Gordon model to condensed matter
physics is that the model is the bosonized version of the fermionic
g-ology or Hubbard models for one-dimensional interacting electron
systems (the Luttinger liquids). In this context, the sine-Gordon
bare parameters $u_{0}$, $K_{0}$ and the non-renormalized velocity
$v$ are connected to the original microscopic couplings defined for
the fermionic models. A comprehensive review on bosonization methods
can be found in the book ``Quantum Physics in One Dimension", by T.
Giamarchi \cite{Giammarchi}.

\subsection{The g-ology model}

The sine-Gordon bare parameters $\tilde{g}_{0}$, $K_{0}$ and $v$ are
related to the 1D g-ology model's microscopic couplings according to
the following expressions \cite{Giammarchi1}
\begin{eqnarray}
\tilde{g}_{0}\rightarrow \tilde{g}_{0\nu}=\left\{
\begin{array}{ll}
0\quad\nu=c\\
\frac{-2g_{1\perp}}{(2\pi\alpha)^{2}}\quad\nu=s\\
\end{array}\right.
\label{g}
\end{eqnarray}

\begin{equation}
K_{0}\rightarrow
K_{0\nu}=\left[\frac{1+y_{4\nu}/2+y_{\nu}/2}{1+y_{4\nu}/2-y_{\nu}/2}\right]^{1/2}
\label{K}
\end{equation}

\begin{equation}
v\rightarrow v_{\nu}=v_{F}[(1+y_{4\nu}/2)^{2}-(y_{\nu}/2)^{2}]^{1/2}
\label{v}
\end{equation}
where
\begin{equation}
y_{\nu}\equiv\frac{g_{\nu}}{\pi v_{F}}
\label{y}
\end{equation}
\begin{equation}
g_{\nu}=g_{1\parallel}-g_{2\parallel}\mp g_{2\perp}
\label{gnu}
\end{equation}
\begin{equation}
g_{4\nu}=g_{4\parallel}\pm g_{4\perp}
\label{gfournu}
\end{equation}
and the sub-indexes $\nu=c,s$ refer, respectively, to the charge and
spin separated sectors of the full bosonized hamiltonian. In Eqs.
(\ref{gnu}) and (\ref{gfournu}), the upper signs refer to $c$ and
the lower ones to $s$.

In the standard g-ology notation, the coupling $g_{4}$ corresponds
to forward scattering between electrons of equal chirality while
$g_{2}$ and $g_{1}$ correspond, respectively, to forward and
backscattering between electrons of different chiralities. Now, the
intensity of each such $g$-scattering may depend on whether the
spins of the two interacting electrons are parallel
($g_{\parallel}$) or anti-parallel ($g_{\perp}$, in lack of a better
notation). Note that for spinless fermions $g_{2}$ and $g_{1}$
processes are identical since one can exchange the outgoing
indiscernible particles. But once the spin comes in the picture,
these two process become intrinsically different and contribute to
the bosonized theory in different ways, as can be seen from the
above equations.

In the general case, when writing models for interacting electrons
one is concerned with the standard Coulomb repulsion between the
particles. In the present context, this translates into positive
$g$-couplings for all processes. However, electrons sometimes can
interact in an attractive way (as for example, through a phonon
mediated coupling). This possibility is taken into account by
allowing for (some) processes with negative g-couplings.

From Eqs. (\ref{g})-(\ref{gfournu}), we see that the Luttinger
liquid separates into a charge sector described by a model of free
bosons with velocity $v_{c}$ and a spin sector that maps into a
bosonic sine-Gordon model with parameters $v_{s}$, $\tilde{g}_{0s}$
and $K_{0s}$. Since $v_{c}\neq v_{s}$ charge and spins excitations
travel independently in the system.

\newpage

From the point of view of the original electronic system, the
massless charge sector represents electrons in a metallic phase. The
behavior of the spin sector is not as simple but can be understood
in the context of the sine-Gordon model phase diagram with bare
parameters determined by Eqs. (\ref{g})-(\ref{gfournu}). This phase
diagram is depicted in Fig. 5 below.
\begin{figure}[h]
\begin{center}
\includegraphics[scale=1.5]{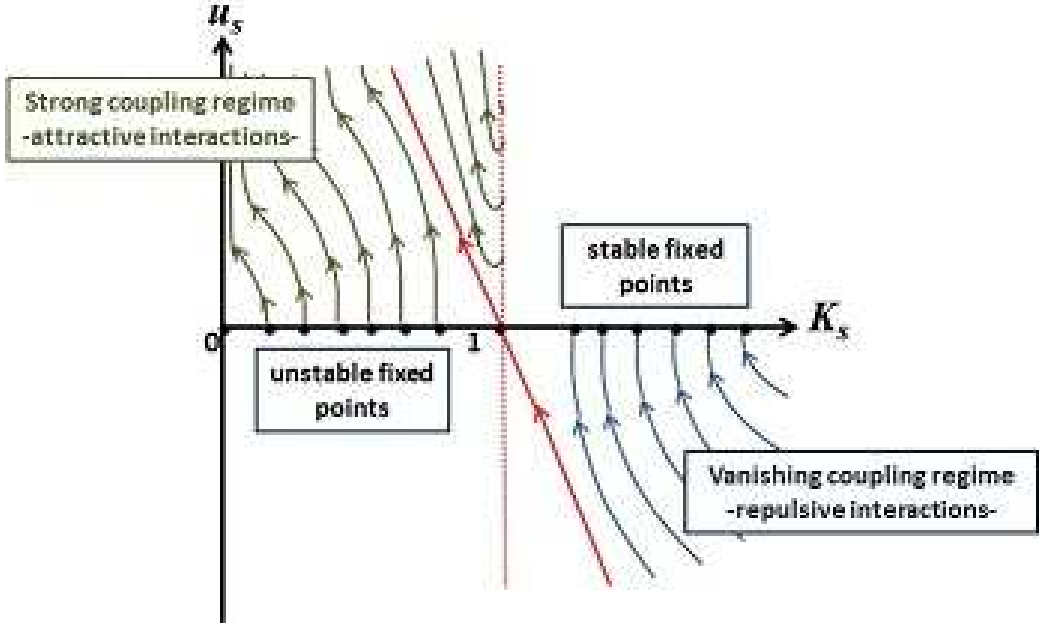}
\end{center}
\caption{Phase diagram for the spin sector of the g-ology model.}
\end{figure}

First of all, note that Eq. (\ref{K}) excludes the $(K<0)$-half of
the full K-T phase diagram. In fact, from Eq. (\ref{Kandu}), a
negative $K$ corresponds to an imaginary $\beta$ which, in turn,
leads to a hyperbolic cosine in Eq. (\ref{H}). Although this is
certainly a mathematical possibility, it is not the case of physical
interest.

Secondly, Eqs. (\ref{K})-(\ref{gnu}) imply that: $K_{0s}>1$ if
$g_{1\parallel}>0$ (for $g_{2}$-processes of comparable intensity),
i.e. if $\tilde{g}_{0s}<0$ (assuming that $g_{1\parallel}$ and
$g_{1\perp}$ have the same sign), and vice-versa. Therefore, for a
repulsive $g_{1}$-interaction, the only physically meaningful region
of the full K-T phase diagram is the one bounded above by the line
$u_{s}=0$ and to the left by the line $K_{s}=1$. On the other hand,
an attractive $g_{1}$-interaction is described by the region bounded
below by the line $u_{s}=0$ and to the right by the line $K_{s}=1$
(and to the left by $K_{s}=0$).

Finally, half of the upper crossover regime of the full K-T phase
diagram was incorporated to the strong coupling regime since, along
the remaining part of the (now relevant) flows, $u_{s}$ increases
monotonically. The lower crossover regime, which would have to be
``artificially" interrupted at the $K_{s}=1$ line, can be excluded
all over based on the argument of weak interactions, i.e. small
$|u_{0s}|$.

The conclusions that can be gleaned from the phase diagram can be
summarized as follows: Repulsive backscattering processes in 1D
electronic systems are irrelevant and the resulting gapless spin
excitations behave, in effect, as a collection of free bosons that
propagate with velocity $v_{s}$ given by Eqs. (\ref{v})-(\ref{gfournu}).
%In this phase, spins cannot sustain long range order.
Attractive backscattering processes flow to the regime of strong
interactions, i.e. are relevant, causing the opening of a gap in the
system's spin sector. A gapped spin excitation means that the spin
$\varphi_{s}$-field gets trapped at a minima of the cosine and
orders, breaking rotational symmetry \footnote{According to the
Mermin-Wagner theorem it is impossible to break a continuous
symmetry in 1+1 dimension, but not a discrete one.}. Assuming that
the nature of electronic interactions, i.e. repulsive or attractive,
is a definite property of a given system, then it is not possible to
drive a phase transition by varying the pair $(K_{0s},u_{0s})$
across the point $(1,0)$.

\newpage

\subsection{The g-ology model at commensurate fillings - umklapp processes}

In 1D electron systems with \emph{commensurate fillings} there is a
fourth type of interaction known as \emph{umpklapp}. The
correspondent coupling constant is termed $g_{3}$ in the g-ology
dictionary. The most known is the case of half-filling that
corresponds to scattering of two left movers to the other side of
the Fermi level through a momentum transfer of $4k_{F}$ from the
lattice. For quarter-filling, an umklapp will be produced by a
similar scattering involving now four particles with a momentum
transfer of $8k_{F}$.
%Such an interaction can only be accessed in higher order
%perturbation theory as it requires three regular two-particles
%scattering processes.

In any case, given that the system is at a commensurate filling, the
bare parameter $\tilde{g}_{0c}$ of Eq. (\ref{g}) is no longer zero
and is associated with a cosine perturbation of the type
\begin{equation}
+\tilde{g}_{0c}\cos(n.\sqrt{8\pi K_{0c}}\varphi-\delta x)
\label{umklapp}
\end{equation}
where
\begin{equation}
\tilde{g}_{0c}=\tilde{g}_{0c}^{n}=\frac{2g_{3,n}}{(2\pi\alpha)^{2}},
\label{gumklapp}
\end{equation}
$n$ is the order of the commensurability (which affects the
amplitude and the wave length of the cosine potential) where $n=1$
corresponds to half-filling, $n=2$ to quarter-filling, etc; and the
parameter $\delta$ measures the deviation (doping) from the
commensurate filling.

The perturbation will oscillate fast due to the phase shift $\delta
x$ and its space integral will vanish unless $\delta x\rightarrow0$.
In other words, away from a commensurate filling (finite $\delta$),
the umklapp is absent and we recover the previous picture of free
bosonic charge excitations. But at a commensurate filling
($\delta=0$), the Luttinger liquid separates into two independent
sine-Gordon models: one for the charge sector with parameters
$v_{c}$, $\tilde{g}_{0c}$ and $\bar{K}_{0c}=n^{2}K_{0c}$ and one for
the spin sector with parameters $v_{s}$, $\tilde{g}_{0s}$ and
$K_{0s}$ (with $v_{\nu}$, $\tilde{g}_{0\nu}$ and $K_{0\nu}$ given by
eqs. (\ref{g})-(\ref{gumklapp})).

Fig. 6 shows the phase diagram for the charge sector of the
$g$-ology model at half-filling ($n=1$), assuming a positive umklapp
coupling $g_{3,n}$. For the charge sector, Eqs.
(\ref{K})-(\ref{gnu}) imply that: $K_{0c}>1$ if
$g_{1\parallel}>g_{2\parallel}+g_{2\perp}$, with no implication on
the value of $\tilde{g}_{0c}$ (that is proportional to $g_{3,n}$
and, thus, positive). The previous condition leads to a number of
possible scenarios. For repulsive interactions, it is verified when
backscattering (between electrons having parallel spins) is more
than twice as intense as forward scattering. The opposite holds for
attractive interactions. If backscattering is repulsive and forward
scattering is attractive, the condition is always verified. In the
opposite scenario, the condition is never verified. In general, it
is possible to drive a phase transition in the charge sector of a 1D
commensurate electronic system by tuning the strength of the
interactions so that the pair $(K_{0c},u_{0c})$ moves across the
separatrix $K_{c}=1+u_{c}/\sqrt{2}$. In the vanishing coupling
regime, the umklapp is irrelevant, charge excitations are gapless
and the system is a metal. In the strong coupling regime, the
umklapp becomes relevant, the charge excitations develop a gap and
the system turns into an insulator. In the crossover regime, the
umklapp is marginal.

Another way to drive a metal-insulator phase transition in a 1D
electronic system is by tuning the filling. Given a fixed
$(K_{0c},u_{0c})$ located in the strong coupling regime, the system
can undergo a metal-insulator phase transition by varying $\delta$,
i.e. the commensurability parameter. This is a phase transition of
incommensurate-commensurate type, also known as Mott-transition.

\begin{figure}[h]
\begin{center}
\includegraphics[scale=1.4]{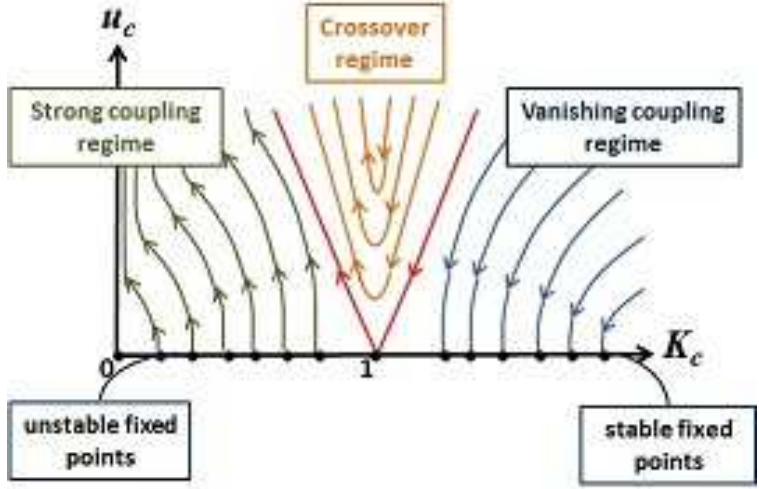}
\end{center}
\caption{Phase diagram for the charge sector of the g-ology model at
half-filling.}
\end{figure}

\newpage

\newpage

\subsection{The Hubbard model}

The sine-Gordon bare parameters $\tilde{g}_{0}$, $K_{0}$ and $v$ are
related to the Hubbard model's microscopic couplings through the
equations\cite{Giammarchi2}
\begin{eqnarray}
\tilde{g}_{0}\rightarrow \tilde{g}_{0\nu}=\left\{
\begin{array}{ll}
0\quad\nu=c\\
\frac{-2U}{(2\pi\alpha)^{2}}\quad\nu=s\\
\end{array}\right.
\label{g1}
\end{eqnarray}

\begin{equation}
vK_{0}\rightarrow v_{\nu}K_{0\nu}=v_{F}
\label{vK}
\end{equation}

\begin{equation}
\frac{v}{K_{0}}\rightarrow\frac{v_{\nu}}{K_{0\nu}}=v_{F}\left(1\pm\frac{U}{\pi v_{F}}\right)
\label{voverK}
\end{equation}
where, as before, $\nu=c,s$ and the upper sign refers to $c$ and the
lower one to $s$. In the Hubbard model, the coupling $U$ represents
an on-site interaction of $g_{1}$ nature with the extra restriction
that, since the interaction is local, it can only take place between
electrons with opposite spins (due to the Pauli principle). The
Hubbard model can be seem as a simplification on the g-ology model.

From Eqs. (\ref{vK}) and (\ref{voverK}) and the fact that the
$K$-parameter is positive, $K_{0s}$ writes in terms of $U$ as:
\begin{equation}
K_{0s}=\frac{1}{\sqrt{1-\frac{U}{\pi v_{F}}}} \label{Ks}
\end{equation}

Also here the Luttinger liquid separates into independent charge and
spin excitations described, respectively, by a free model and a
sine-Gordon model with their respective parameters. From Eq.
(\ref{Ks}), $U>0\Rightarrow K_{0s}>1$ and vice-versa. Therefore, for
a system with a repulsive and weak enough Hubbard interaction, the
pair of bare parameters $(K_{0s},u_{0s})$ lies inside the $u_{s}<0$
vanishing coupling, irrelevant, regime of the full K-T phase
diagram. In this case, only the gapless phase is accessible for the
spin system which consists of free bosonic excitations. Meanwhile,
for an attractive $U$, the pair of bare parameters $(K_{0s},u_{0s})$
will fall into either the $u_{s}>0$ strong coupling regime or in the
left half of the crossover regime where the interaction becomes
relevant. In either cases, the spin sector develops a gap and the
spin field orders. The spin sector's phase diagram is the same as in
Fig. 5 obtained in the context of for the $g$-ology model. Here,
again, we do not expect a phase transition between the gapless and
the gapped phases of the spin excitations developing in a metal
where the nature of the Hubbard on-site interaction is either
repulsive or attractive.

If the system is at half-filling, the charge sector develops an
umklapp interaction of the form
\begin{equation}
+\tilde{g}_{0c}\cos(\sqrt{8\pi K_{0c}}\varphi-\delta x)
\label{umklapp2}
\end{equation}
where in the Hubbard model language:
\begin{equation}
\tilde{g}_{0c}=\frac{2U}{(2\pi\alpha)^{2}}, \label{gumklapp1}
\end{equation}

For commensurate fillings other than $1/2$, the umklapp interaction
assumes similar expressions.

From Eqs. (\ref{vK}) and (\ref{voverK}), $K_{0c}$ is given in terms
of $U$ as:
\begin{equation}
K_{0c}=\frac{1}{\sqrt{1+\frac{U}{\pi v_{F}}}} \label{Kc}
\end{equation}

Thus, at commensurate fillings, the Luttinger liquid separates into
two independent sine-Gordon models: one for the charge and one for
the spin sector with their correspondent parameters. Fig. 7 shows
the phase diagram for the charge sector of the Hubbard model at
half-filling.

\vspace{.7cm}

\begin{figure}[h]
\begin{center}
\includegraphics[scale=1.5]{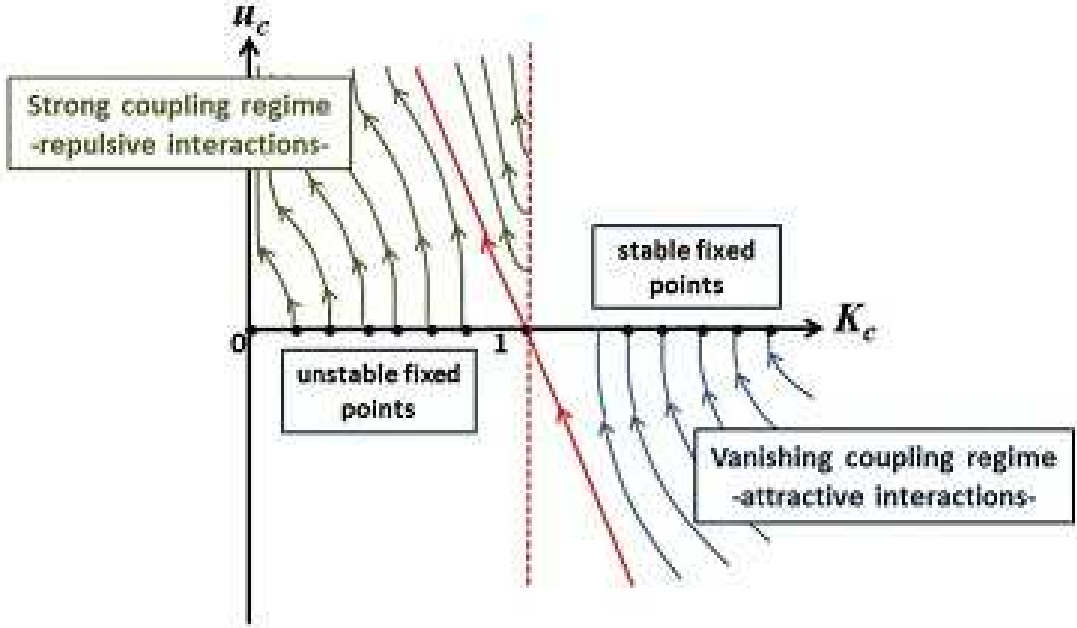}
\end{center}
\caption{Phase diagram for the charge sector of the Hubbard model at
half-filling.}
\end{figure}

\newpage

From Eq. (\ref{Kc}), $U>0\Rightarrow K_{0c}<1$ and vice-versa.
Therefore, for a repulsive interaction, the pair of bare parameters
$(K_{0c},u_{0c})$ falls inside either the $u_{c}>0$ strong coupling
regime or in the left half of the crossover regime. In both
situations the interaction is relevant and the system opens up a
gap, becoming an insulator. On the other hand, a weak enough
attractive interaction puts $(K_{0c},u_{0c})$ inside the $u_{c}<0$
vanishing coupling regime where the interaction is irrelevant. In
this regime, the gapless charge excitations remain in the metallic
phase. As before, one cannot drive a metal-insulator phase
transition between the repulsive and attractive portions of the
phase diagram in a system where the interactions have a definite
nature.

In summary, the Hubbard model describes the following types of 1D
systems of interacting electrons: Away from commensurability, the
system is a metal described by gapless charge excitations and, if
the Hubbard interaction is repulsive, gapless spin excitations that
preserve rotational symmetry, or gapped and symmetry breaking spin
excitations if the interaction is attractive. For commensurate
fillings, the system will be an insulator formed of gapped charge
excitations and gapless spin excitations for a repulsive
interaction, while an attractive interaction leads to a metal with
gapless charge excitations and gapped spin excitations.

%$\star$ Estimate for the gaps\\

%In section IV.B, we provided the expression for the gap as a
%function of the general sine-Gordon bare parameters $K_{0}$ and
%$u_{0}$: $\Delta=c_{0}|u_{0}|^{1/2(1-K_{0})}$. In the current
%scenario, we would like to estimate the charge/spin gap
%$\Delta_{c/s}$ that might arise in the charge/spin sector of a 1D
%electron system in terms of the physical fermionic couplings. To do
%that we just substitute into the previous expression for the gap the
%formulae established in this chapter connecting the sine-Gordon bare
%parameters $K_{0c/s}$ and $u_{0c/s}$ and the fermionic couplings.
%Finally, one should have in mind that this estimate is only good
%deep inside the insulator/paramagnetic regime of the charge/spin
%sector's K-T phase diagram.

\end{document}